\documentclass[preprint,showpacs,preprintnumbers,amsmath,amssymb,aps]{revtex4}

\usepackage{graphicx}
\usepackage{dcolumn}
\usepackage{bm}

\begin{document}

\title{Derivation of the Gauge Link in Light Cone Gauge}

\author{Jian-Hua Gao}
\email{gaojh79@ustc.edu.cn}
\affiliation{\mbox{Department of Modern Physics, University of Science and Technology of China,}\\
Hefei, Anhui 230026, China}
\affiliation{Department of Physics, Shandong University, Jinan, Shandong 250100, China}

\date{\today}


\begin{abstract}
In light cone gauge, a  gauge link at light cone infinity is
necessary  for transverse momentum-dependent parton distribution to
restore the gauge invariance  in some specific boundary conditions.
We derive such transverse gauge link in a more regular and general
method. We find the gauge link at light cone infinity naturally
arises from the contribution of the pinched poles:
 one is from the quark propagator and
the other is hidden in the gauge vector field in light cone gauge. Actually, in the
amplitude level, we have obtained a more general gauge link  over the hypersurface
at light cone infinity which is beyond the transverse direction. The  difference of such gauge link
between semi-inclusive deep inelastic scattering and Drell-Yan processes can also be obtained directly and
clearly in our derivation.
\end{abstract}


\maketitle


\section{introduction}
Nucleon structure functions are  physical observables  and can be
measured in deep inelastic scattering (DIS). In the naive parton
model\cite{Fey71}, the structure functions are expressed in terms of
the probability of finding quarks and gluons in the parent nucleon.
In collinear QCD factorization formulas, such structure functions
can be given by compact operator matrix elements of the
target\cite{ColSop82}

\begin{equation}
\label{CollinearPDFs}
q(x)=\frac{1}{2}
\int \frac{d y^-}{2\pi}{\rm e}^{- i x p^{+}y^-}
\langle P |\bar\psi (y^-,\vec{0}_\perp) n\!\!\!\slash \mathcal{L}[y^-,\vec{0}_\perp; 0,\vec{0}_\perp] \psi (0,\vec{0}_\perp)| P \rangle\, ,
\end{equation}
where
\begin{equation}
\label{CollinearGL}
\mathcal{L}[y^-,\vec{0}_\perp; 0,\vec{0}_\perp]\equiv
P \exp \left(- i g \int^{y^-}_0 d \xi^{-} A^+ ( \xi^-,\vec{0}_\perp )
\right) \, ,
\end{equation}
is the gauge link between the quark fields, which arises from final
state interactions between the struck quark and the target
spectators. In  Eq.~(\ref{CollinearPDFs}) and
Eq.~(\ref{CollinearGL}), all fields are evaluated at equal $y^+=0$.
Since structure functions, as physical observables, should not be
dependent on the gauge that we choose, it  is necessary to introduce
such gauge link to ensure the gauge invariance of matrix element. In
the light cone gauge $A^+=0$, where the path-ordered exponential in
Eq.~(\ref{CollinearGL}) reduces to unity,  we can identify the quark
distribution in Eq.~(\ref{CollinearPDFs}) as a probability
distribution as we made in naive parton model. Actually, in
collinear structure function such as Eq.~(\ref{CollinearPDFs}), we
can always select a clever gauge to vanish the gauge link. But when
we consider the transverse-momentum  dependent quark distribution,
such naive manipulation will result in inconsistency. In the
nonsingular gauge, in which the gauge potential vanishes at the
space-time infinity, the transverse-momentum parton distribution is
defined in the literature as
\cite{ColSop82,Collins:1989bt,Collins:2003fm}
\begin{eqnarray}
\label{TMD}
q(x,\vec{k}_\perp)&=&\frac{1}{2}
\int \frac{d y^-}{2\pi}\frac{d^2 \vec{y}_\perp}{(2\pi)^2}{\rm e}^{- i x p^{+}y^{-}+i\vec{k}_\perp\cdot\vec{y}_\perp}\nonumber\\
&&\times\langle P |\bar\psi (y^-,\vec{y}_\perp) n\!\!\!\slash \mathcal{L}^\dagger[\infty,\vec{y}_\perp;y^-,\vec{y}_\perp ]
 \mathcal{L}[\infty,\vec{0}_\perp;0,\vec{0}_\perp ]\psi (0,\vec{0}_\perp)| P \rangle\, ,
\end{eqnarray}
where
\begin{equation}
\label{TMDGL}
\mathcal{L}[\infty,\vec{y}_\perp; y^-,\vec{y}_\perp]\equiv
P \exp \left(- i g \int_{y^-}^\infty d \xi^{-} A^+ ( \xi^-, \vec{y}_\perp)
\right) \, ,
\end{equation}
and  all fields are evaluated at equal $y^+=0$. From Lorentz invariance, parity invariance and
time reversal invariance, the transverse-momentum parton distribution can be decomposed into the following expressions,
\begin{eqnarray}
\label{decompose}
q(x,\vec{k}_\perp)&=& f(x,k_\perp)+\vec{S}\cdot(\hat{\vec{p}}\times \vec{k}_\perp)f_{1T}^\perp(x,k_\perp)/M
\end{eqnarray}
where $\vec{S}$ is the spin of the target nucleon and
$\hat{\vec{p}}\ $ is a unit vector along the direction of the target
momentum in infinite momentum frame. The function $f_{1T}^\perp(x,k_\perp)$ is
just the Sivers function and  can contribute to single spin
asymmetries. It is verified in Ref.\cite{Collins:1992kk} that the Sivers
function  vanishes unless there is the gauge link in Eq.(\ref{TMDGL}),
which is yielded by the final state interactions \cite{Brodsky:2002cx}.
 In the light cone gauge, however, it seems as if the gauge link in
Eq.~(\ref{TMDGL}) would become unity  and the final interaction
vanish accordingly too. Hence there will be inconsistent results
from different gauges, which is impossible, since physical observables
should not depend on the gauge by choice.  Ji and Yuan in
\cite{Ji:2002aa} have shown that the final state interaction effects
in single spin asymmetry can be recovered properly in the light cone
gauge by taking into account a transverse gauge link at
$y^-=+\infty$. Further in \cite{Belitsky:2002sm}, Belitsky, Ji, and
Yuan demonstrate the existence of extra leading twist contributions
from transverse components of the gauge potential at the light cone
infinity. It turns out that these contributions just form a
transverse gauge link in light cone gauge.  In this paper, we will
give another more regular and systematic method to obtain such
transverse gauge link in light cone gauge.
We find the gauge link at light cone infinity will arise naturally from the pinched poles, one of which is provided by the quark propagator and
the other is hidden in the gauge vector field in light cone gauge.
Actually, it turn out that
we  obtain a more general gauge link over  hypersurface $y^-=\infty$, instead of
only transverse gauge link. The difference of such  gauge link
between semi-inclusive deep inelastic scattering (SIDIS) and Drell-Yan (DY) processes can also be shown directly
and clearly in our derivation.
The paper is organized as follows: in the next section, we will introduce
some kinetics definitions and notations which will be involved all
through our paper. In Sec.III, we would like to give a brief review on the
singularity in light cone gauge and different prescriptions for different
light cone pole structures. Then in Sec.IV,  we will devote to deriving the
 gauge link in light cone gauge in SIDIS process. In Sec.V,
 we will deal with the DY process and compare it with the SIDIS process. A very short summary is given in the end.
 Other relevant work on the transverse gauge link can be found in the literature \cite{Boer:2003cm,Idilbi:2008vm}.

\section{some definitions and notations}
In  studying SIDIS or DY process, it is convenient to  choose  the light cone coordinate system
in which we  introduce two lightlike vectors $n^{\mu}$ and $\bar{n}^{\mu}$,

\begin{equation}
n^{\mu}=(0,1,\vec{0}_\perp)\ ,\ \ \bar{n}^{\mu}=(1,0,\vec{0}_\perp)\ ,\ \ n\cdot\bar{n}=1\ .
\end{equation}
With these basis vectors, we may write any  vector $k^{\mu}$ as $(k^{+},k^{-},\vec{k}_{\perp})$, where $k^{+}=k\cdot n,\ k^{-}=k\cdot \bar{n}$.
For example, in SIDIS process, we  choose the proton  infinite momentum frame, in which the proton's momentum and
the virtual photon's momentum are given by, respectively,
\begin{equation}
p^{\mu}=p^{+}\bar{n}^{\mu},\ \ \ q^{\mu}=-x_B p^{\mu}+\frac{Q^2}{2x_B p^{+}}{n}^{\mu}.
\end{equation}
where $x_B=Q^2/2p\cdot q$ and $Q^2=-q^2$.

In order to make the  derivation  more compact and elegant in the following sections, let us introduce some notations.
For any momentum vector $k^{\mu}$ and the gauge potential vector $A^{\mu}$,
we will manipulate the following decomposition:
\begin{eqnarray}
k^{\mu}&=&{\tilde{k}}^{\mu}+xp^{\mu},\ \ A^{\mu}={\tilde{A}}^{\mu}+ A^{+} \bar{n}^{\mu}
\end{eqnarray}
where ${\tilde{k}}^{\mu}=(0,k^{-},\vec{k}_{\perp})$, $x=k^+/p^+$, and $\tilde{A}^{\mu}=(0,A^{-},\vec{A}_{\perp})$.
For any coordinate vector $y^{\mu}$ , we will make the dual decomposition,
\begin{eqnarray}
y^{\mu}&=&{\dot{y}}^{\mu}+y^{-} n^{\mu}
\end{eqnarray}
where $\dot{y}^{\mu}=(y^{+},0,\vec{y}_{\perp})$. When there is no confusion, we will rewrite $y^\mu$ as $(y^-,\dot{y})$. With such notations,
we have $k\cdot y=\tilde{k}\cdot\dot{y} +xp^{+}y^{-}$,
and in light cone gauge where $A^+=0$ ,we  also have $ A^{\mu}={\tilde{A}}^{\mu}$. It should be noted that in light cone coordinate,
the covariant vector and contravariant vector are related by $A^+=A_-$, $A^-=A_+$ and $A^{\perp}=-A_{\perp}$.

\section{spurious singularity in light cone gauge }
The light cone gauge $n\cdot A=0$
is widely used in perturbative QCD calculations \cite{Dokshitzer:1978hw,Mueller:1981sg}, and under such
a physical gauge condition,  the probability interpretation is expected to hold. The Yang-Mills theories, quantized in light cone
gauge, have been studied by several authors \cite{Bassetto:1984dq,Leibbrandt:1983pj}. However, when we calculate with
the gauge propagator in such gauge in perturbation theory, we have to introduce some spurious pole to regularize associated
light cone singularity. There have been a variety of  prescriptions  suggested to handle the singularities
{\cite{Bassetto:1983bz,Lepage:1980fj,Leibbrandt:1987qv,Slavnov:1987yh,Kovchegov:1997pc}, in which most
attempts were pragmatic. The literature \cite{Belitsky:2002sm,Bassetto:1981gp} states that in general, in light cone gauge, the
gauge potential can not be arbitrarily set to vanish at the infinity, the spurious singularities, characteristic of all the axial gauges,
are physically related to the boundary conditions that one can impose on the potentials at the infinity.
In our paper, we will  consider three different boundary conditions as in \cite{Belitsky:2002sm}, i.e.
\begin{eqnarray}
\label{poles}
&&\textrm{Advanced} : \ \ \ \ \ \ \ \ \ \tilde{A}(\infty,\dot{y})=0 \nonumber\\
&&\textrm{Retarded} : \ \ \ \ \ \ \ \ \ \tilde{A}(-\infty,\dot{y})=0 \nonumber\\
&&\textrm{Antisymmetric} :\ \ \tilde{A}(-\infty,\dot{y})+\tilde{A}(\infty,\dot{y})=0.
\end{eqnarray}
The typical integration  we will meet with in our derivation is the Fourier transformation of the gauge potential such as,
\begin{eqnarray}
\tilde{\cal{A}}_\rho(k^+,\dot{y})\equiv\int_{-\infty}^{\infty}dy^{-}\textrm{e}^{ik^{+}y^{-}}\tilde{A}_{\rho}(y^{-},\dot{y})
\end{eqnarray}
Manipulating this integration by parts, we  obtain
\begin{eqnarray}
\label{integration}
\int_{-\infty}^{\infty}dy^{-}\textrm{e}^{ik^{+}y^{-}}\tilde{A}_{\rho}(y^{-},\dot{y})
=[\frac{i}{k^{+}}]\int_{-\infty}^{\infty}dy^{-}\textrm{e}^{ik^{+}y^{-}}\partial^{+}\tilde{A}_{\rho}(y^{-},\dot{y})
\end{eqnarray}
where $\partial^{+}=\partial_-={\partial}/{\partial y^-}$. Since the  boundary condition is set, the term $[\frac{i}{k^{+}}]$ can be regularized by
definite prescription,
\begin{eqnarray}
\label{regularization}
&&\textrm{Advanced} : \ \ \ \ \ \ \ \ \ [\frac{i}{k^{+}}]=\frac{i}{k^{+}-i\epsilon} \nonumber\\
&&\textrm{Retarded} : \ \ \ \ \ \ \ \ \ [\frac{i}{k^{+}}]=\frac{i}{k^{+}+i\epsilon} \nonumber\\
&&\textrm{Antisymmetric} :\ \ [\frac{i}{k^{+}}]=\frac{1}{2}(\frac{i}{k^{+}+i\epsilon}+\frac{i}{k^{+}-i\epsilon}).
\end{eqnarray}
where the last propose is just the conventional principal value regulation when the antisymmetry boundary condition
is assigned. Hence, we notice  that there is a secret pole structure in gauge potential in momentum space. We will
show that it is just this pole that will contribute to the final gauge link at the light cone infinity.

The easiest way to illustrate the validity of such regularization is just to set
\begin{equation}
\label{stepfunction}
\tilde{A}_{\rho}(y^{-})=
\left\{
\begin{array}{ll}
{\rm Advanced:} &  \theta(-y^{-})\\
{\rm Retarted:} &  \theta(y^{-})\\
{\rm Antysymmetry:}  & \frac{1}{2}\left[\theta(y^{-})-\theta(-y^{-})\right]
\end{array}
\right.
\end{equation}
where the function $\theta(y^{-})$ is the usual step function. It is  a trivial exercise to show that they can result in the proper pole structure as
we present in Eq.\ (\ref{poles}).

As we mentioned above, in the light cone gauge, we can not impose on the gauge potential   the boundary condition
both $\tilde{A}_{\rho}(+\infty,\dot{y})=0$ and $\tilde{A}_{\rho}(-\infty,\dot{y})=0$. We can only choose either of them as the boundary
condition to remove the residual gauge freedom and the other one will be subjected to satisfy the field equation or the request that
the total gauge energy momentum is finite. However, as a matter of fact, we can still impose a weaker condition, that the gauge potential
must be a pure gauge.  In the Abelian case,
\begin{equation}
\label{abelian}
\tilde{A}_{\rho}(\pm\infty,\dot{y})=\tilde{\partial}_{\rho}\phi(\pm\infty,\dot{y})
\end{equation}
or in the non-Abelian case
\begin{equation}
\label{nonabelian}
\tilde{A}_{\rho}(\pm\infty,\dot{y})=\omega^{-1}(\pm\infty,\dot{y})\tilde{\partial}_{\rho}\omega(\pm\infty,\dot{y})
\end{equation}
where $\omega=\textrm{exp}(i\phi)$.
In the non-Abelian case, $\tilde{A}_{\rho}\equiv \tilde{A}^a_{\rho}t^a$ and $\phi \equiv \phi^a t^a$ where $t^a$ are the generators of non-Abelian group in the
fundamental representation. Keeping the leading term in the Tailor expansion of $\omega$ around $\phi$, we  recover the same expression as
 Eq.~(\ref{abelian}) in the Abelian case.
It follows that
\begin{equation}
\label{pathintegral}
\phi(+\infty,\dot{y})=-\int_{\dot{y}}^{\dot{\infty}}d\dot{\xi}\cdot\tilde{A}(+\infty,\dot{\xi})
\end{equation}
where the  integral runs over any  path on the hypersurface $y^-=\infty$.
Notice that this equation  always holds for Abelian gauge potential, and holds for the non-Abelian case only when the $\phi $ is small. It will be interesting
thing to investigate what the nonleading terms contribute to in the non-Abelian case, which is beyond the scope of  this paper. We will show that
the linear term, such as in Eq.~({\ref{pathintegral}}) will lead to the  gauge link at the light cone infinity.

\section{Gauge link in light cone gauge in SIDIS}

\begin{figure}
\begin{center}
\includegraphics[scale=0.9]
{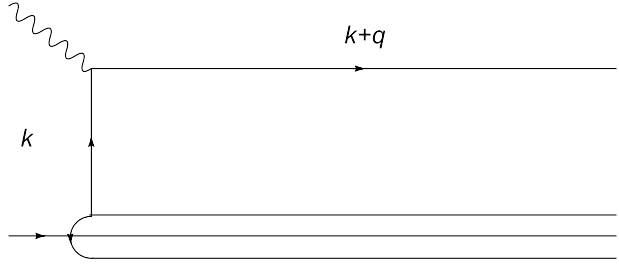} \caption{ The tree diagram in DIS process}
\label{TGL0}
\end{center}
\end{figure}

In DIS process, the hadronic tensor is defined by

\begin{equation}
\label{DIStensor}
W^{\mu\nu} = \frac{1}{4 \pi} \sum_{X} \int \frac{d^3 p_J}{(2 \pi)^3}
(2 \pi)^4 \delta^{(4)} \left( P_{_X} + p_J - p - q \right)
\langle P | j^\mu (0) | p_J, X \rangle
\langle p_J, X | j^\nu (0) | P \rangle \, .
\end{equation}

The tree scattering amplitude corresponding to Fig.\ \ref{TGL0} reads
\begin{equation}
M_{0}^\mu =\langle p_J, X | j^\mu (0) | P \rangle_{(0)}
=\bar{u}(k+q)\gamma^\mu \langle X|\psi(0)|P\rangle\,,
\end{equation}
where $k$ denotes the momentum of intial quark scattered by the photon with momentum $q$.

\begin{figure}
\begin{center}
\includegraphics[scale=0.9]
{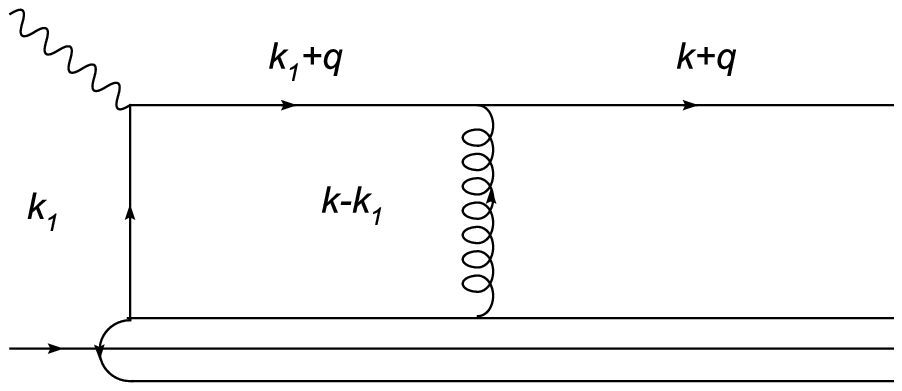} \caption{The one-gluon exchange diagram in DIS process}
\label{TGL1}
\end{center}
\end{figure}
The one-gluon amplitude in light cone gauge corresponding to Fig.\ \ref{TGL1} reads,

\begin{eqnarray}
M_{1}^\mu &=&\int \frac{d^4k_{1}}{(2\pi)^4}\int d^4y_{1}\ e^{i(k-k_{1})\cdot y_{1}}\nonumber\\
&&\times \bar{u}(k+q)
\gamma^{\rho_{1}}\frac{{k\!\!\!\slash}_{1}\!\!+q\!\!\!\slash}{(k_{1}+q)^2+i\epsilon}
\langle X|\tilde{A}_{\rho_{1}}(y_{1})\gamma^{\mu}\psi(0)|P\rangle\,.
\end{eqnarray}
The quark propagator can be decomposed into two parts,

\begin{equation}
\frac{{k\!\!\!\slash}_{1}\!\!+q\!\!\!\slash}{(k_{1}+q)^2+i\epsilon}=\frac{1}{2p\cdot(\hat{k}_{1}+q)}
\left[\frac{{\hat{k}\!\!\!\slash}_{1}\!\!+q\!\!\!\slash}{(x_{1}-\hat{x}_{1}+i\epsilon)}+p\!\!\!\slash\right]\,,
\label{decompose}
\end{equation}
where $\hat{k}_1 \equiv(\hat{x}_1p^+,k_1^-,k_{1\perp})$ with $\ \hat{x}_1=\hat{k}^{+}/p^{+}=x_{B}+k_{\perp}^2/2p\cdot (k_1+q)\ $
is determined by the on-shell condition $\ (\hat{k}_1+q)^2=0\ $.
 Actually,
to obtain the Eq.~(\ref{decompose}), we have neglected the contribution
\begin{equation}
\frac{\Theta(-k_1^- -q^-)}{2p\cdot(\hat{k}_{1}+q)}
\left[\frac{{\hat{k}\!\!\!\slash}_{1}\!\!+q\!\!\!\slash}{(x_{1}-\hat{x}_{1}-i\epsilon)}+p\!\!\!\slash\right]
\end{equation}
which will contribute at higher twist level  since they  vanish in the limit $q^-\rightarrow +\infty$.
The last
term in Eq.~(\ref{decompose}) is the so-called ``contact'' term of a normal propagator which does not propagate
along the light cone coordinate \cite{Qiu:1988dn}. Such a contact term will always result in higher twist contribution
and does not contribute to  gauge link at all. Hence, when we are considering the leading twist contribution in our following derivation,
we can just drop such contact terms and only keep the pole terms, i.e. the first term in Eq.\ (\ref{decompose}):

\begin{eqnarray}
\hat{M}_{1}^\mu &=&\int \frac{d^3\tilde{k}_{1}}{(2\pi)^4}\int d^3\dot{y}_{1}\ \int \frac{p^{+}dx_{1}}{2\pi}\int dy_{1}^{-}
\ e^{i(\tilde{k}-\tilde{k}_{1})\cdot \dot{y}_{1}+i({x}-{x}_{1})p^{+}y^{-}}\nonumber\\
&&\times\frac{1}{2p\cdot(\hat{k}_{1}+q)} \bar{u}(\hat{k}+q)
\gamma^{\rho_{1}}\frac{{\hat{k}\!\!\!\slash}_{1}\!\!+q\!\!\!\slash}{(x_{1}-\hat{x}_{1}+i\epsilon)}
\langle X|\tilde{A}_{\rho_{1}}(y_{1})\gamma^{\mu}\psi(0)|P\rangle
\end{eqnarray}
where  another notation $\hat{M}_{1}^\mu$ with an extra $\hat{\ }$ is introduced to remind us that the only pole term
is kept, and we have also separate the integral over $x_1$ and $y^-_1$ from the others which means we will
finish integrating them out first in the following. Before proceeding further, we should first choose a specific boundary condition for the gauge potential $\tilde{A}_{\rho}$ at
infinity. Let us start with the retarded boundary condition $\tilde{A}(\infty,\dot{y})=0 $. Using the Eq.\ (\ref{integration}) accordingly
which corresponds to retarded boundary condition, we have
\begin{eqnarray}
\label{M1hat1}
\hat{M}_{1}^\mu &=&\int \frac{d^3\tilde{k}_{1}}{(2\pi)^4}\int d^3\dot{y}_{1}\ \int \frac{dx_{1}}{2\pi}\int dy_{1}^{-}
\ e^{i(\tilde{k}-\tilde{k}_{1})\cdot \dot{y}_{1}+i({x}-{x}_{1})p^{+}y^{-}}\nonumber\\
&&\times \bar{u}({k}+q)
\gamma^{\rho_{1}}\frac{{\hat{k}\!\!\!\slash}_{1}\!\!+q\!\!\!\slash}{2p\cdot(\hat{k}_{1}+q)}
\frac{1}{(x_{1}-\hat{x}_{1}+i\epsilon)}\frac{i}{({x}-{x}_{1}+i\epsilon)} \nonumber\\
&&\times\langle X|\partial^{+}\tilde{A}_{\rho_{1}}(y_{1})\gamma^{\mu}\psi(0)|P\rangle\,.
\end{eqnarray}
Now we can finish  integrating over $x_1$ and $y_1^{-}$ first,
\begin{eqnarray}
\label{contour1}
&&\int \frac{dx_{1}}{2\pi}\int dy_{1}^{-}e^{i({x}-{x}_{1})p^{+}y^{-}}
\frac{1}{(x_{1}-\hat{x}_{1}+i\epsilon)}\frac{i}{(x-{x}_{1}+i\epsilon)}
\partial^{+}\tilde{A}_{\rho_{1}}(y_{1}) \nonumber\\
&=&\frac{1}{{x}-\hat{x}_{1}}\int dy_{1}^{-}
\left(\theta(y^{-})e^{i({x}-\hat{x}_{1})p^{+}y^{-}}+\theta(-y^{-})\right)\partial^{+}\tilde{A}_{\rho_{1}}(y_{1})\nonumber\\
&=&\frac{1}{{x}-\hat{x}_{1}}\int dy_{1}^{-}
\left(\theta(y^{-})+\theta(-y^{-})\right)\partial^{+}\tilde{A}_{\rho_{1}}(y_{1})+ \textrm{higher twist}\nonumber\\
&=&\frac{1}{{x}-\hat{x}_{1}}\tilde{A}_{\rho_{1}}(+\infty,\dot{y}_{1})+ \textrm{higher twist}\,,
\end{eqnarray}
where only the leading term in the Tailor expansion of the phase factor $e^{i({x}-\hat{x}_{1})p^{+}y^{-}}$
is kept, because the other terms are proportional to $({x}-\hat{x}_{1})^n=\left[k_{\perp}^2/2p\cdot (k+q)-k_{1\perp}^2/2p\cdot (k_1+q)\right]^n$ $(n\ge 1)$,
which will contribute at higher twist level.  Only keep leading twist contribution and inserting Eq.\ (\ref{contour1}) into Eq.\ (\ref{M1hat1}), we  have

\begin{eqnarray}
\hat{M}_{1}&=&\int \frac{d^3\tilde{k}_{1}}{(2\pi)^4}\int d^3\dot{y}_{1}\
\ e^{i(\tilde{k}-\tilde{k}_{1})\cdot \dot{y}_{1}}\nonumber\\
&&\times \bar{u}({k}+q)
\gamma^{\rho_{1}}\frac{{\hat{k}\!\!\!\slash}_{1}\!\!+q\!\!\!\slash}{2p\cdot (\hat{k}_{1}+q)}
\frac{1}{{x}-\hat{x}_{1}}\langle X|\tilde{A}_{\rho_{1}}(+\infty,\dot{y}_{1})\psi(0)|P\rangle\,.
\end{eqnarray}
Using Eq.\ ({\ref{abelian}}) and performing the integration by parts over $\dot{y}_{1}$ where
$\tilde{\partial}_{\rho}\to -i(\tilde{k}-\tilde{k}_{1})_\rho$, we  obtain
\begin{eqnarray}
\hat{M}_{1}&=&\int \frac{d^3\tilde{k}_{1}}{(2\pi)^4}\int d^3\dot{y}_{1}\
\ e^{i(\tilde{k}-\tilde{k}_{1})\cdot \dot{y}_{1}}\nonumber\\
&&\times \bar{u}({k}+q)
(\tilde{k}\!\!\!\slash-\tilde{k}\!\!\!\slash_{1})\frac{{\hat{k}\!\!\!\slash}_{1}\!\!+q\!\!\!\slash}{2p\cdot (\hat{k}_{1}+q)}
\frac{-i}{{x}-\hat{x}_{1}}\langle X|\phi(+\infty,\dot{y}_{1})\psi(0)|P\rangle\,.
\end{eqnarray}
To carry out the matrix algebra further, we note that
\begin{eqnarray}
\label{algebra1}
\tilde{k}\!\!\!\slash-\tilde{k}\!\!\!\slash_{1}
=({k}\!\!\!\slash+q\!\!\!\slash)-(\hat{k}\!\!\!\slash_{1}+q\!\!\!\slash)-({x}-\hat{x}_{1})p\!\!\!\slash\,,
\end{eqnarray}
together with the on-shell conditions
\begin{eqnarray}
\label{algebra2}
\bar{u}({k}+q)({k}\!\!\!\slash+q\!\!\!\slash)=0,\textrm{and}\ (\hat{k}\!\!\!\slash_{1}+q\!\!\!\slash)^2=0\,.
\end{eqnarray}
Using these equations, we  reduce the $\hat{M}_{1}$ into
\begin{eqnarray}
\label{M1last2}
\hat{M}_{1}&=&\int \frac{d^3\tilde{k}_{1}}{(2\pi)^4}\int d^3\dot{y}_{1}\
\ e^{i(\tilde{k}-\tilde{k}_{1})\cdot \dot{y}_{1}}\nonumber\\
&&\times \bar{u}({k}+q)
p\!\!\!\slash(\hat{k}\!\!\!\slash_{1}\!\!+q\!\!\!\slash)\frac{-i}{2p\cdot (\hat{k}_{1}+q)}
\langle X|\phi(+\infty,\dot{y}_{1})\psi(0)|P\rangle \nonumber\\
&=&\bar{u}({k}+q)
\langle X|i\phi(+\infty,{0})\psi(0)|P\rangle \nonumber\\
&&+\int \frac{d^3\tilde{k}_{1}}{(2\pi)^4}\int d^3\dot{y}_{1}\
\ e^{i(\tilde{k}_{2}-\tilde{k}_{1})\cdot \dot{y}_{1}}\nonumber\\
&&\times \bar{u}({k}+q)(\tilde{k}\!\!\!\slash-\tilde{k}\!\!\!\slash_{1})
p\!\!\!\slash\frac{-i}{2p\cdot (\hat{k}_{1}+q)}
\langle X|\phi(+\infty,\dot{y}_{1})\psi(0)|P\rangle\,.
\end{eqnarray}
Since the last term in Eq.\ (\ref{M1last2}) only contribute to higher twist, keeping only the leading twist contribution,
we  finally obtain,
\begin{eqnarray}
\label{M1last}
\hat{M}_{1}
&=&\bar{u}({k}+q)
\langle X|i\phi(+\infty,{0})\psi(0)|P\rangle\,.
\end{eqnarray}

So far, the previous derivations  have been restricted  to the retarded boundary condition where $\tilde{A}(-\infty,\dot{y})=0$, now let
us turn to the other two boundary conditions.  When we assign the advanced boundary condition $\tilde{A}(+\infty,\dot{y})=0$, which means that
 we should choose the advanced one in Eq.\ (\ref{integration}) and Eq.\ (\ref{regularization}). Such a sign change in the pole structure
will lead to replacing the integration in Eq.\ (\ref{contour1}) by,

\begin{eqnarray}
\label{contour2}
&&\int \frac{dx_{1}}{2\pi}\int dy_{1}^{-}e^{i({x}-{x}_{1})p^{+}y^{-}}
\frac{1}{(x_{1}-\hat{x}_{1}+i\epsilon)}\frac{i}{(x-{x}_{1}-i\epsilon)}
\partial^{+}\tilde{A}_{\rho_{1}}(y_{1}) \nonumber\\
&=&\frac{1}{{x}-\hat{x}_{1}}\int dy_{1}^{-}
\left(\theta(y^{-})e^{i({x}-\hat{x}_{1})p^{+}y^{-}}-\theta(y^{-})\right)\partial^{+}\tilde{A}_{\rho_{1}}(y_{1})\nonumber\\
&=& \textrm{higher twist}\,.
\end{eqnarray}
We note that, different from retarded case,  the leading contributions from two poles have canceled  each other completely, and there will be no
gauge link at all. As shown by \cite{Belitsky:2002sm}, all final state interactions have been included into the initial
state light cone wave functions.

If we choose the antisymmetry boundary condition, which corresponds to the principal value regularization, we  have
\begin{eqnarray}
\label{contour3}
&&\int \frac{dx_{1}}{2\pi}\int dy_{1}^{-}e^{i({x}-{x}_{1})p^{+}y^{-}}
\frac{1}{(x_{1}-\hat{x}_{1}+i\epsilon)}\textrm{PV}\frac{i}{(x-{x}_{1})}
\partial^{+}\tilde{A}_{\rho_{1}}(y_{1}) \nonumber\\
&=&\frac{1}{{x}-\hat{x}_{1}}\int dy_{1}^{-}
\frac{1}{2}\left(2\theta(y^{-})e^{i({x}-\hat{x}_{1})p^{+}y^{-}}-\theta(y^{-})+\theta(-y^{-})\right)\partial^{+}\tilde{A}_{\rho_{1}}(y_{1})\nonumber\\
&=&\frac{1}{{x}-\hat{x}_{1}}\int dy_{1}^{-}
\frac{1}{2}\left(\theta(y^{-})+\theta(-y^{-})\right)\partial^{+}\tilde{A}_{\rho_{1}}(y_{1})+ \textrm{higher twist}\nonumber\\
&=&\frac{1}{{x}-\hat{x}_{1}}\tilde{A}_{\rho_{1}}(+\infty,\dot{y}_{1})+ \textrm{higher twist}\,,
\end{eqnarray}
where $\mbox{PV}$ denotes principal value.
The above result appear  the same as the one in the retarded boundary condition.  The difference between retarded and principal value regularization is that  final state scattering effects appear
only through  the  gauge link in principal regularization, while they appear through  both the gauge link and initial light cone wave
functions in retarded regularization. Such detailed discussion and  illustration can be found in Ref.\cite{Belitsky:2002sm}.
In the above derivation, we notice that the pinched poles are
needed to pick up the gauge potential at the light cone infinity, which will be shown to result in the gauge link that we expect.
In the following, we will only concentrate on the retarded boundary condition in the following derivation.
\begin{figure}
\begin{center}
\includegraphics[scale=0.9]
{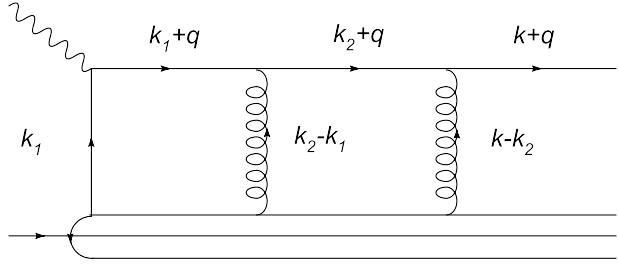} \caption{The two-gluon exchange diagram in DIS process} \label{TGL2}
\end{center}
\end{figure}

Now let us consider further the two-gluon exchange  scattering amplitude in Fig. \ref{TGL2},
\begin{eqnarray}
M_{2}^\mu &=&\int \frac{d^4k_{2}}{(2\pi)^4}\frac{d^4k_{1}}{(2\pi)^4}\int d^4y_{2}d^4y_{1}
\ e^{i(k-k_{2})\cdot y_{2}+i(k_{2}-k_{1})\cdot y_{1}}\nonumber\\
&&\times \bar{u}(k+q)
\gamma^{\rho_{2}}\frac{{k\!\!\!\slash}_{2}\!\!+q\!\!\!\slash}{(k_{2}+q)^2+i\epsilon}
\gamma^{\rho_{1}}\frac{{k\!\!\!\slash}_{1}\!\!+q\!\!\!\slash}{(k_{1}+q)^2+i\epsilon}
\langle X|\tilde{A}_{\rho_{2}}(y_{2})\tilde{A}_{\rho_{1}}(y_{1})\gamma^\mu\psi(0)|P\rangle\,.
\end{eqnarray}
Just following what we did in the $M_{1}^\mu$, we drop the contact terms which do not contribute in leading twist level and
label the residual terms as $\hat{M}_{2}$, which is given by
\begin{eqnarray}
\hat{M}_{2}^\mu&=&\int \frac{d^3\tilde{k}_{2}}{(2\pi)^3}\frac{d^3\tilde{k}_{1}}{(2\pi)^3}\int d^3\dot{y}_{2}d^3\dot{y}_{1}
\ \int \frac{p^{+}dx_{2}}{2\pi}\frac{p^{+}dx_{1}}{2\pi}\int dy_{2}^{-}dy_{1}^{-}\nonumber\\
&&\times e^{i(\tilde{k}-\tilde{k}_{2})\cdot \dot{y}_{2}+i(\tilde{k}_{2}-\tilde{k}_{1})\cdot \dot{y}_{1}
+i({x}-{x}_{2})p^{+}y_{2}^{-}+i({x}_{2}-{x}_{1})p^{+}y_{1}^{-}}\nonumber\\
&&\times \bar{u}(k+q)
\gamma^{\rho_{2}}\frac{{\hat{k}\!\!\!\slash}_{2}\!\!+q\!\!\!\slash}{2p\cdot(\hat{k}_{2}+q)}
\gamma^{\rho_{1}}\frac{{\hat{k}\!\!\!\slash}_{1}\!\!+q\!\!\!\slash}{2p\cdot(\hat{k}_{1}+q)}
\frac{1}{(x_{2}-\hat{x}_{2}+i\epsilon)}\frac{1}{(x_{1}-\hat{x}_{1}+i\epsilon)}\nonumber\\
&&\times \langle X|\tilde{A}_{\rho_{2}}(y_{2})\tilde{A}_{\rho_{1}}(y_{1})\gamma^\mu\psi(0)|P\rangle\,.
\end{eqnarray}
Still with the help of the regularization in Eqs.~(\ref{integration}) and (\ref{regularization}), let us do  integrating
over $x_1$ and $y^{-}_1$ first,
\begin{eqnarray}
\hat{M}_{2}^\mu&=&\int \frac{d^3\tilde{k}_{2}}{(2\pi)^3}\frac{d^3\tilde{k}_{1}}{(2\pi)^3}\int d^3\dot{y}_{2}d^3\dot{y}_{1}
\ \int \frac{dx_{2}}{2\pi}\frac{p^{+}dx_{1}}{2\pi}\int dy_{2}^{-}dy_{1}^{-}\nonumber\\
&&\times e^{i(\tilde{k}-\tilde{k}_{2})\cdot \dot{y}_{2}+i(\tilde{k}_{2}-\tilde{k}_{1})\cdot \dot{y}_{1}
+i({x}-{x}_{2})p^{+}y_{2}^{-}+i({x}_{2}-{x}_{1})p^{+}y_{1}^{-}}\nonumber\\
&&\times \bar{u}(k+q)
\gamma^{\rho_{2}}\frac{{\hat{k}\!\!\!\slash}_{2}\!\!+q\!\!\!\slash}{2p\cdot(\hat{k}_{2}+q)}
\gamma^{\rho_{1}}\frac{{\hat{k}\!\!\!\slash}_{1}\!\!+q\!\!\!\slash}{2p\cdot(\hat{k}_{1}+q)}
\frac{1}{(x_{2}-\hat{x}_{2}+i\epsilon)}\frac{i}{(\hat{x}-{x}_{2}+i\epsilon)}
\frac{1}{(x_{1}-\hat{x}_{1}+i\epsilon)} \nonumber\\
&&\times\langle X|\partial^{+}\tilde{A}_{\rho_{2}}(y_{2})\tilde{A}_{\rho_{1}}(y_{1})\gamma^\mu\psi(0)|P\rangle \nonumber\\
&=&\int \frac{d^3\tilde{k}_{2}}{(2\pi)^3}\frac{d^3\tilde{k}_{1}}{(2\pi)^3}\int d^3\dot{y}_{2}d^3\dot{y}_{1}
\ \int \frac{p^{+}dx_{1}}{2\pi}\int dy_{1}^{-}
\  e^{i(\tilde{k}-\tilde{k}_{2})\cdot \dot{y}_{2}+i(\tilde{k}_{2}-\tilde{k}_{1})\cdot \dot{y}_{1}
+i({x}-{x}_{1})p^{+}y_{1}^{-}}\nonumber\\
&&\times \bar{u}(k+q)
\gamma^{\rho_{2}}\frac{{\hat{k}\!\!\!\slash}_{2}\!\!+q\!\!\!\slash}{2p\cdot(\hat{k}_{2}+q)}
\gamma^{\rho_{1}}\frac{{\hat{k}\!\!\!\slash}_{1}\!\!+q\!\!\!\slash}{2p\cdot(\hat{k}_{1}+q)}
\frac{1}{(x-\hat{x}_{2}+i\epsilon)}\frac{1}{(x_{1}-\hat{x}_{1}+i\epsilon)} \nonumber\\
&&\times\langle X|\tilde{A}_{\rho_{2}}(+\infty,\dot{y}_{2})\tilde{A}_{\rho_{1}}(y_{1})\gamma^\mu\psi(0)|P\rangle\,.
\end{eqnarray}
Further integrating over $x_2$ and $y^{-}_2$, which is totally the same as what we did with $x_1$ and $y^{-}_1$.
The results read
\begin{eqnarray}
\hat{M}_{2}^\mu
&=&\int \frac{d^3\tilde{k}_{2}}{(2\pi)^3}\frac{d^3\tilde{k}_{1}}{(2\pi)^3}\int d^3\dot{y}_{2}d^3\dot{y}_{1}
\  e^{i(\tilde{k}-\tilde{k}_{2})\cdot \dot{y}_{2}+i(\tilde{k}_{2}-\tilde{k}_{1})\cdot \dot{y}_{1}}\nonumber\\
&&\times \bar{u}(k+q)
\gamma^{\rho_{2}}\frac{{\hat{k}\!\!\!\slash}_{2}\!\!+q\!\!\!\slash}{2p\cdot(\hat{k}_{2}+q)}
\gamma^{\rho_{1}}\frac{{\hat{k}\!\!\!\slash}_{1}\!\!+q\!\!\!\slash}{2p\cdot(\hat{k}_{1}+q)}
\frac{1}{(x-\hat{x}_{2}+i\epsilon)}\frac{1}{(x-\hat{x}_{1}+i\epsilon)} \nonumber\\
&&\times\langle X|\tilde{A}_{\rho_{2}}(+\infty,\dot{y}_{2})\tilde{A}_{\rho_{1}}(+\infty,\dot{y}_{1})\gamma^\mu\psi(0)|P\rangle \nonumber\\
&=&\int \frac{d^3\tilde{k}_{2}}{(2\pi)^3}\frac{d^3\tilde{k}_{1}}{(2\pi)^3}\int d^3\dot{y}_{2}d^3\dot{y}_{1}
\  e^{i(\tilde{k}-\tilde{k}_{2})\cdot \dot{y}_{2}+i(\tilde{k}_{2}-\tilde{k}_{1})\cdot \dot{y}_{1}}\nonumber\\
&&\times \bar{u}(k+q)
\gamma^{\rho_{2}}\frac{{\hat{k}\!\!\!\slash}_{2}\!\!+q\!\!\!\slash}{2p\cdot(\hat{k}_{2}+q)}
\gamma^{\rho_{1}}\frac{{\hat{k}\!\!\!\slash}_{1}\!\!+q\!\!\!\slash}{2p\cdot(\hat{k}_{1}+q)}
\frac{1}{(x-\hat{x}_{2}+i\epsilon)}\frac{1}{(x-\hat{x}_{1}+i\epsilon)} \nonumber\\
&&\times\langle X|\tilde{\partial}_{\rho_{2}}\phi(+\infty,\dot{y}_{2})\tilde{\partial}_{\rho_{1}}\phi(+\infty,\dot{y}_{1})\gamma^\mu\psi(0)|P\rangle\,.
\end{eqnarray}
Now we are in a position to perform integrating over $\tilde{k}_2$ and $\dot{y}_2$. Thanks to the integration by parts and the algebras
given in Eq.\ (\ref{algebra1}) and Eq.\ (\ref{algebra2}), we  obtain

\begin{eqnarray}
\hat{M}_{2}^\mu
&=&\int \frac{d^3\tilde{k}_{1}}{(2\pi)^3}\int d^3\dot{y}_{1}
\  e^{i(\tilde{k}-\tilde{k}_{1})\cdot \dot{y}_{1}}
 \bar{u}(k+q)
\gamma^{\rho_{1}}\frac{{\hat{k}\!\!\!\slash}_{1}\!\!+q\!\!\!\slash}{2p\cdot(\hat{k}_{1}+q)}
\frac{1}{(x-\hat{x}_{1}+i\epsilon)} \nonumber\\
&&\times \langle X|i\phi(+\infty,\dot{y}_{1})\tilde{\partial}_{\rho_{1}}\phi(+\infty,\dot{y}_{1})\gamma^\mu\psi(0)|P\rangle \nonumber\\
&=&\int \frac{d^3\tilde{k}_{1}}{(2\pi)^3}\int d^3\dot{y}_{1}
\  e^{i(\tilde{k}_{2}-\tilde{k}_{1})\cdot \dot{y}_{1}}
 \bar{u}(k+q)
\gamma^{\rho_{1}}\frac{{\hat{k}\!\!\!\slash}_{1}\!\!+q\!\!\!\slash}{2p\cdot(\hat{k}_{1}+q)}
\frac{1}{(x-\hat{x}_{1}+i\epsilon)} \nonumber\\
&&\times \langle X|\frac{i}{2}\tilde{\partial}_{\rho_{1}}\phi^2(+\infty,\dot{y}_{1})\gamma^\mu\psi(0)|P\rangle
\end{eqnarray}
Repeat what we did with $\tilde{k}_2$ and $\dot{y}_2$ above, and we can finish integrating over
$\tilde{k}_1$ and $\dot{y}_1$ and finally arrive
\begin{eqnarray}
\hat{M}_{2}^\mu&=& \bar{u}(k+q)
\langle X|\frac{i^2}{2!}\phi^2(+\infty,{0})\gamma^\mu\psi(0)|P\rangle
\end{eqnarray}
All through calculating  $\hat{M}_{2}$, as we did with $\hat{M}_{1}$, we have neglected the higher twist contributions and only keep
the leading twist terms.
From ${M}_{1}$ to ${M}_{2}$, it is obvious that our procedure  can be easily extended to $n$-gluon exchange
amplitude $M_{n}$ in Fig. \ref{TGLn}, which is given by

\begin{figure}
\begin{center}
\includegraphics[scale=0.9]
{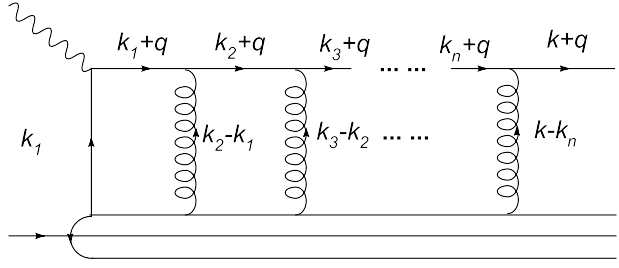} \caption{The $n$-gluon exchange diagram in DIS process} \label{TGLn}
\end{center}
\end{figure}

\begin{eqnarray}
\hat{M}_{n}
&=&
\int\prod_{j=1}^{n}\frac{d^3{\tilde{k}_{j}}}{(2\pi)^3}d^3\dot{y}_{j}
e^{i(\tilde{k}_{n}-\tilde{k}_{n-1}) \cdot\dot{y}_{n}+i(\tilde{k}_{n-1}-\tilde{k}_{n-2})\cdot\dot{y}_{n-1}+\ ...\ ...\ +i(\tilde{k}_{2}-\tilde{k}_{1})\cdot\dot{y}_{2}}\nonumber\\
&&\times
\prod_{j=1}^{n}
\frac{p^{+}d{x}_{j}}{2\pi}dy^{-}_{j}
e^{i({x}_{n+1}-{x}_{n})p^{+} {y}_{n}^{-}+i({x}_{n}-{x}_{n-1})p^{+}{y}_{n-1}^{-}+\ ...\ ...\ +i({x}_{2}-{x}_{1})p^{+}{y}_{1}^{-}}\nonumber\\
&&\times \bar{u}({k}+q)
\gamma^{\rho_{n}}\frac{{\hat{k}\!\!\!\slash}_{n}\!\!+q\!\!\!\slash}{2p\cdot(\hat{k}_{n}+q)}
\ \ ...\ \ ...\ \
\gamma^{\rho_{1}}\frac{{\hat{k}\!\!\!\slash}_{1}\!\!+q\!\!\!\slash}{2p\cdot(\hat{k}_{1}+q)} \nonumber\\
&&\times\frac{1}{(x_{n}-\hat{x}_{n}+i\epsilon)}\ \ ...\ \ ...\ \
\frac{1}{(x_{1}-\hat{x}_{1}+i\epsilon)}  \nonumber\\
&&\times\langle X|\tilde{A}_{\rho_{n}}(y_{n})\tilde{A}_{\rho_{n-1}}(y_{n-1})\ ...\ ...\ \tilde{A}_{\rho_{1}}(y_{1})\psi(0)|P\rangle\,.
\end{eqnarray}
We  first finish integrating from $x_{n},y^{-}_{n}$ to $x_{1},y^{-}_{1}$ one by one. Keeping the leading twist
contribution, we have,
\begin{eqnarray}
\hat{M}_{n}
&=&
\int\prod_{j=1}^{n}\frac{d^3{\tilde{k}_{j}}}{(2\pi)^3}d^3\dot{y}_{j}
e^{i(\tilde{k}_{n+1}-\tilde{k}_{n}) \cdot\dot{y}_{n}+i(\tilde{k}_{n}-\tilde{k}_{n-1})\cdot\dot{y}_{n-1}+\ ...\ ...\ +i(\tilde{k}_{2}-\tilde{k}_{1})\cdot\dot{y}_{2}}\nonumber\\
&&\times \bar{u}({k}+q)
\gamma^{\rho_{n}}\frac{{\hat{k}\!\!\!\slash}_{n}\!\!+q\!\!\!\slash}{2p\cdot(\hat{k}_{n}+q)}
\ \ ...\ \ ...\ \
\gamma^{\rho_{1}}\frac{{\hat{k}\!\!\!\slash}_{1}\!\!+q\!\!\!\slash}{2p\cdot(\hat{k}_{1}+q)} \nonumber\\
&&\times \frac{1}{(\hat{x}_{n+1}-\hat{x}_{n})}\ \ ...\ \ ...\ \
\frac{1}{(\hat{x}_{2}-\hat{x}_{1})} \nonumber\\
&&\times \langle X|\tilde{A}_{\rho_{n}}(+\infty,\dot{y}_{n})\tilde{A}_{\rho_{n-1}}(+\infty,\dot{y}_{n-1})
\ ...\ ...\ \tilde{A}_{\rho_{1}}(+\infty,\dot{y}_{1})\psi(0)|P\rangle\,,
\end{eqnarray}
 or using
\begin{equation}
\tilde{A}_{\rho_{n}}(+\infty,\dot{y}_{n})=\tilde{\partial}_{\rho_{n}}\phi(+\infty,\dot{y}_{n})\,,
\end{equation}
we  rewrite it as
\begin{eqnarray}
\hat{M}_{n}
&=&
\int\prod_{j=1}^{n}\frac{d^3{\tilde{k}_{j}}}{(2\pi)^3}d^3\dot{y}_{j}
e^{i(\tilde{k}_{n+1}-\tilde{k}_{n}) \cdot\dot{y}_{n}+i(\tilde{k}_{n}-\tilde{k}_{n-1})\cdot\dot{y}_{n-1}+\ ...\ ...\ +i(\tilde{k}_{2}-\tilde{k}_{1})\cdot\dot{y}_{2}}\nonumber\\
&&\times \bar{u}({k}+q)
\gamma^{\rho_{n}}\frac{{\hat{k}\!\!\!\slash}_{n}\!\!+q\!\!\!\slash}{2p\cdot(\hat{k}_{n}+q)}
\ \ ...\ \ ...\ \
\gamma^{\rho_{1}}\frac{{\hat{k}\!\!\!\slash}_{1}\!\!+q\!\!\!\slash}{2p\cdot(\hat{k}_{1}+q)} \nonumber\\
&&\times \frac{1}{(\hat{x}_{n+1}-\hat{x}_{n})}\ \ ...\ \ ...\ \
\frac{1}{(\hat{x}_{2}-\hat{x}_{1})} \nonumber\\
&&\times \langle X|\tilde{\partial}_{\rho_{n}}\phi(+\infty,\dot{y}_{n})\tilde{\partial}_{\rho_{n-1}}\phi(+\infty,\dot{y}_{n-1})
\ ...\ ...\  \tilde{\partial}_{\rho_{1}}\phi(+\infty,\dot{y}_{1})\psi(0)|P\rangle\,.
\end{eqnarray}
Continue to integrating over from $\tilde{k}_n$ and $\dot{y}_n$ to $\tilde{k}_1$ and $\dot{y}_1$ one by one, we can finally have
\begin{eqnarray}
\hat{M}_{n}
&=&
\bar{u}({k}+q)\langle X|\frac{i^n}{n!}\phi^n(+\infty,0)\psi(0)|P\rangle\,.
\end{eqnarray}
As a final step,  we should resum to all orders and obtain
\begin{eqnarray}
\sum_{n=0}^{\infty}\hat{M}_{n}
&=&\bar{u}({k}+q)\langle X|\textrm{exp}\left(i\phi(+\infty,0)\right)\psi(0)|P\rangle\,,
\end{eqnarray}
or the  more conventional form
\begin{eqnarray}
\label{final-dis}
\sum_{n=0}^{\infty}\hat{M}_{n}
&=&\bar{u}({k}+q)\langle X|P\textrm{exp}\left(-i\int_{\dot{y}}^{\dot{\infty}}
d\dot{\xi}\cdot\tilde{A}(+\infty,\dot{\xi})\right)\psi(0)|P\rangle\,,
\end{eqnarray}
where $P\textrm{exp}\left(-i\int_{\dot{y}}^{\dot{\infty}}d\dot{\xi}\cdot\tilde{A}(+\infty,\dot{\xi})\right)$ is just the gauge link
that we tried to derive. It should be noted that the gauge link we obtain in the final result Eq.(\ref{final-dis})
is  over the hypersurface at light cone infinity along any path integral, not restricted along the transverse direction, which means that it is more general
than what Belitsky, Ji and Yuan have obtained in Ref.\cite{Belitsky:2002sm}.

\section{Gauge link in light cone gauge in DY}
\begin{figure}
\begin{center}
\includegraphics[scale=0.9]
{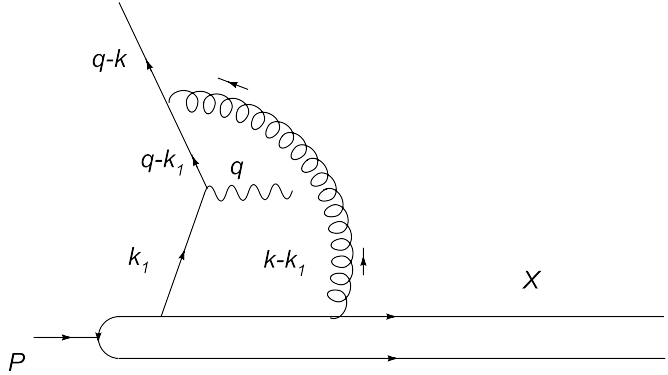} \caption{The one-gluon exchange diagram in DY process} \label{Drell-Yan}
\end{center}
\end{figure}

Now let us turn to the DY process, which is represented in Fig.\ (\ref{Drell-Yan}), where, for brevity, we have fixed
the target to be  a nucleon and the projectile to be just an antiquark, $q$ is the virtual photon's momentum
and $q-k$ and $p$ is momentum of the projectile and target respectively. Such simplifying does not lose any generality
when we are only considering how to derive the gauge link,  but  it will be more convenient and
manifest to  compare with the SIDIS process.
 We still choose the light cone coordinate system, and
use the two lightlike vectors $n^\mu$ and $\bar{n}^\mu$ to fix ``plus'' and ``minus'' directions. All the notations  and conventions are
the same as in DIS. The one-gluon exchange amplitude reads

\begin{eqnarray}
M_{1(DY)}^\mu &=&\int \frac{d^4k_{1}}{(2\pi)^4}\int d^4y_{1}\ e^{i(k-k_{1})\cdot y_{1}}\nonumber\\
&&\times \bar{u}(q-k)\gamma^{\rho_{1}}\frac{{q\!\!\!\slash-k\!\!\!\slash}_{1}\!\!}{(q-k_{1})^2+i\epsilon}
\langle X|\tilde{A}_{\rho_{1}}(y_{1})\gamma^{\mu}\psi(0)|P\rangle\,.
\end{eqnarray}
Dropping the contact terms and  assigning  the retarded boundary condition, we  rewrite it as
\begin{eqnarray}
\label{M1hat1DY}
\hat{M}_{1(DY)}^\mu &=&\int \frac{d^3\tilde{k}_{1}}{(2\pi)^4}\int d^3\dot{y}_{1}\ \int \frac{dx_{1}}{2\pi}\int dy_{1}^{-}
\ e^{i(\tilde{k}-\tilde{k}_{1})\cdot \dot{y}_{1}}
e^{i({x}-{x}_{1})p^{+}y^{-}}\nonumber\\
&&\times \bar{u}(q-k)\gamma^{\rho_{1}}\frac{{q\!\!\!\slash-\hat{k}\!\!\!\slash}_{1}\!\!}{2p\cdot(\hat{k}_{1}-q)}
\frac{1}{(x_{1}-\hat{x}_{1}-i\epsilon)}\frac{i}{({x}-{x}_{1}+i\epsilon)} \nonumber\\
&&\times\langle X|\partial^{+}\tilde{A}_{\rho_{1}}(y_{1})\gamma^{\mu}\psi(0)|P\rangle\,.
\end{eqnarray}
It should be noticed  the difference of the pole structure between Eq.(\ref{M1hat1DY}) and Eq.(\ref{M1hat1}).
 Just like we did in the SIDIS,  we can finish  integrating over $x_1$ and $y_1^{-}$ first,
\begin{eqnarray}
\label{contour4}
&&\int \frac{dx_{1}}{2\pi}\int dy_{1}^{-}e^{i({x}-{x}_{1})p^{+}y^{-}}
\frac{1}{(x_{1}-\hat{x}_{1}-i\epsilon)}\frac{i}{(x-{x}_{1}+i\epsilon)}
\partial^{+}\tilde{A}_{\rho_{1}}(y_{1}) \nonumber\\
&=&-\frac{1}{{x}-\hat{x}_{1}}\int dy_{1}^{-}
\left(\theta(-y^{-})e^{i({x}-\hat{x}_{1})p^{+}y^{-}}-\theta(-y^{-})\right)\partial^{+}\tilde{A}_{\rho_{1}}(y_{1})\nonumber\\
&=&-\frac{1}{{x}-\hat{x}_{1}}\int dy_{1}^{-}
\left(\theta(-y^{-})-\theta(-y^{-})\right)\partial^{+}\tilde{A}_{\rho_{1}}(y_{1})+ \textrm{higher twist}\nonumber\\
&=& \textrm{higher twist}\,.
\end{eqnarray}
Opposite to the case in SIDIS, the retarded boundary condition does not lead to
the  gauge link  at the light cone infinity  and hence all the final state interaction effects must be shifted into the initial light cone wave
functions. For the advanced boundary condition, we  have
\begin{eqnarray}
\label{M1hat1DY-A}
\hat{M}_{1(DY)}^\mu &=&\int \frac{d^3\tilde{k}_{1}}{(2\pi)^4}\int d^3\dot{y}_{1}\ \int \frac{dx_{1}}{2\pi}\int dy_{1}^{-}
\ e^{i(\tilde{k}-\tilde{k}_{1})\cdot \dot{y}_{1}}
e^{i({x}-{x}_{1})p^{+}y^{-}}\nonumber\\
&&\times \bar{u}(q-k)\gamma^{\rho_{1}}\frac{{q\!\!\!\slash-\hat{k}\!\!\!\slash}_{1}\!\!}{2p\cdot(\hat{k}_{1}-q)}
\frac{1}{(x_{1}-\hat{x}_{1}-i\epsilon)}\frac{i}{({x}-{x}_{1}-i\epsilon)} \nonumber\\
&&\times\langle X|\partial^{+}\tilde{A}_{\rho_{1}}(y_{1})\gamma^{\mu}\psi(0)|P\rangle\,.
\end{eqnarray}
Finish  integrating over $x_1$ and $y_1^{-}$:
\begin{eqnarray}
\label{contour5}
&&\int \frac{dx_{1}}{2\pi}\int dy_{1}^{-}e^{i({x}-{x}_{1})p^{+}y^{-}}
\frac{1}{(x_{1}-\hat{x}_{1}-i\epsilon)}\frac{i}{(x-{x}_{1}-i\epsilon)}
\partial^{+}\tilde{A}_{\rho_{1}}(y_{1}) \nonumber\\
&=&-\frac{1}{{x}-\hat{x}_{1}}\int dy_{1}^{-}
\left(\theta(-y^{-})e^{i({x}-\hat{x}_{1})p^{+}y^{-}}+\theta(y^{-})\right)\partial^{+}\tilde{A}_{\rho_{1}}(y_{1})\nonumber\\
&=&-\frac{1}{{x}-\hat{x}_{1}}\int dy_{1}^{-}
\left(\theta(-y^{-})+\theta(y^{-})\right)\partial^{+}\tilde{A}_{\rho_{1}}(y_{1})+ \textrm{higher twist}\nonumber\\
&=&\frac{1}{{x}-\hat{x}_{1}}\tilde{A}_{\rho_{1}}(-\infty,\dot{y}_{1})+ \textrm{higher twist}\,.
\end{eqnarray}
If we choose the antisymmetry boundary condition, we have,
\begin{eqnarray}
\label{contour3}
&&\int \frac{dx_{1}}{2\pi}\int dy_{1}^{-}e^{i({x}-{x}_{1})p^{+}y^{-}}
\frac{1}{(x_{1}-\hat{x}_{1}-i\epsilon)}\textrm{PV}\frac{i}{(x-{x}_{1})}
\partial^{+}\tilde{A}_{\rho_{1}}(y_{1}) \nonumber\\
&=&\frac{1}{{x}-\hat{x}_{1}}\int dy_{1}^{-}
\frac{1}{2}\left(-2\theta(-y^{-})e^{i({x}-\hat{x}_{1})p^{+}y^{-}}-\theta(y^{-})+\theta(-y^{-})\right)\partial^{+}\tilde{A}_{\rho_{1}}(y_{1})\nonumber\\
&=&-\frac{1}{{x}-\hat{x}_{1}}\int dy_{1}^{-}
\frac{1}{2}\left(\theta(y^{-})+\theta(-y^{-})\right)\partial^{+}\tilde{A}_{\rho_{1}}(y_{1})+ \textrm{higher twist}\nonumber\\
&=&\frac{1}{{x}-\hat{x}_{1}}\tilde{A}_{\rho_{1}}(-\infty,\dot{y}_{1})+ \textrm{higher twist}\,.
\end{eqnarray}
The difference between advanced and principal value regularization in the DY process is that  final state scattering effects appear
only through  the  gauge link in principal value regularization, while they appear through  both the gauge link and initial light cone wave
functions in advanced regularization.
It follows that,
\begin{eqnarray}
\hat{M}_{1(DY)}
&=&\bar{u}(q-{k})\langle X|\left(-i\int^{\dot{y}}_{-\dot{\infty}}d\dot{\xi}\cdot\tilde{A}(-\infty,\dot{\xi})\right)\psi(0)|P\rangle\,.
\end{eqnarray}
Just as in the SIDIS process, the pinched poles are indispensable to produce a finite contribution in the leading
twist level.
Following the same line
as the one carried out in SIDIS, we  show that the gauge link in advanced or antisymmetry boundary conditions  is given by
\begin{eqnarray}
\sum_{n=0}^{\infty}\hat{M}_{n(DY)}
&=&\bar{u}(q-{k})\langle X|P\textrm{exp}\left(-i\int^{\dot{y}}_{-\dot{\infty}}d\dot{\xi}\cdot\tilde{A}(-\infty,\dot{\xi})\right)\psi(0)|P\rangle\,.
\end{eqnarray}
It should be noted that
the light cone infinity $y^-=+\infty$ has been replaced by $y^-=-\infty$, reflecting that the gauge link arises from the initial state
interactions rather than from the final state.

To summarize, in light cone gauge, we should choose a specific boundary condition first to fix the residual gauge freedom. Using the proper
regularization corresponding to specific boundary condition, we can obtain the residual gauge link at infinity along the light cone coordinate.
We find the gauge link at light cone infinity arises naturally from the pinched poles: one is from the quark propagator and
the other is hidden in the gauge vector field in light cone gauge.
Actually, it turns out that
we  obtain a more general gauge link over  hypersurface $y^-=\infty$, which is beyond the transverse gauge link. The difference of such  gauge link
between SIDIS and DY processes can also be obtained directly
and clearly in our derivation. We expect our regularization method will also be valuable to make it possible to
perform higher twist calculations in light cone gauge more unambiguously.

\begin{acknowledgments}
I would like to thank Jian Zhou for helpful discussions. This work
 was supported, in part, by the National Natural Science Foundation of China
under the Grant No. 10525523,  the Department of Science and Technology of
Shandong Province and  the China Postdoctoral Science Foundation
funded project under Contract No. 20090460736.

\end{acknowledgments}

\end{document}